\documentclass{article}
\usepackage{LaThuileFPSpro}
\begin{document}
\title{ 
A NEW PERSPECTIVE ON PARTON DISTRIBUTIONS IN NUCLEI
  }
\author{
  Simonetta Liuti        \\
  {\em University of Virginia, Charlottesville, Virginia, U.S.A.} \\
  }
\maketitle

\baselineskip=11.6pt

\begin{abstract}
We present recent progress on the study of the deep inelastic structure of 
nuclei that improves our current understanding of the 
mechanisms of nuclear modifications of parton distribution functions.      
\end{abstract}
\newpage
\section{Introduction}
The determination of the quark and gluon structure of nuclei 
remains an important open, and currently unsolved question in Quantum 
Chromodynamics (QCD).
Its detailed knowledge is, in fact, fundamental 
for understanding the occurrence of a deconfined quark-gluon phase 
both in the high baryon density regime present in
neutron stars, and at high temperatures accessible both at the Relativistic 
Heavy Ion Collider (RHIC), and at the 
Large Hadron Collider (LHC).  
Moreover, the theoretical uncertainty related to nuclear effects hinders 
the extraction of possible contributions of new physics 
from a number of high precision experiments. This uncertaintly will 
be particularly felt in the forthcoming
generation of experiments using neutrino beams 
\cite{Kulagin:2006dg}.
   
Theoretical efforts have so far concentrated on two apparently distinct areas.
Many studies were dedicated on one side to the interpretation of 
Deep Inelastic Scattering (DIS) experiments.
In the past twenty years it has been established that sizable $A$-dependent effects 
affect nuclear DIS cross sections (for e recent review see \cite{Deshpande:2005wd}), 
thus suggesting that the nuclear parton distributions
could not be described as a collection
of ``quasi-free'' nucleons 
whose partonic structure would 
remain unaffected by the relatively weak nuclear forces.  
On the other side, a number of recent 
exclusive electron-nucleus scattering experiments
allow for investigations of the nucleon form factors for bound nucleons. 
Their initial outcome is also suggestive of non trivial deformations
of the charge and magnetic current distributions of bound nucleons
\cite{Strauch:2002wu}. 

Recently, a more comprehensive object, the Generalized Parton Distribution (GPD) was 
introduced that interpolates between the Parton Distribution Functions (PDFs) from DIS, 
and the form factors (for reviews see \cite{Diehl:2003ny,Belitsky:2005qn}). GPDs are the soft components 
in the hadronic tensor for Deeply Virtual Compton Scattering (DVCS). A relation 
of GPDs to Wigner distributions was also recently uncovered in Ref.\cite{Belitsky:2003nz} after the initial observation
that they carry information on the location of partons inside the hadron \cite{Burkardt:2004bv}. 
GPDs provide a unique tool to explore the spatial distributions of 
quarks and gluons in nuclei.


%
%
\section{Nuclear Effects in Deep Inelastic Scattering: Status and Perspectives}
\begin{figure}[t]
  \vspace{6.0cm}
  \includegraphics{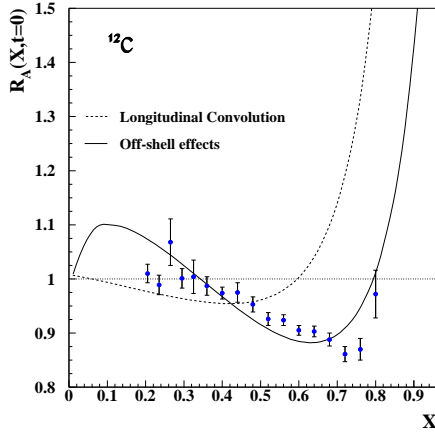}
  \caption{\it Example of nuclear effects from a forward ($t=0$) DIS experiment. Only data at large 
$x$ are displayed 
Ref.\protect\cite{Gomez:1993ri}. The full curve is our current description of large $x$ nuclear effects, including the 
effect of transverse degrees of freedom explained in the text. The dashed curve displayed for
comparison, includes conventional nuclear degrees of freedom only. 
}
    \label{fig1}
\end{figure}

The study of nuclear modifications of DIS type cross sections with respect to
the free nucleon ones started in the mid 80's with the discovery by 
the EMC collaboration of up to 20\% discrepancies 
in the iron structure function $F_2$, at values Bjorken $x \approx 0.5$.
It is currently understood that there are four distinct regions 
characterized by nuclear effects of different nature: $x<0.1$, 
the {\it shadowing} region, $0.1<x<0.2$, the {\it anti-shadowing} region, 
$0.2 <x < 0.6$ the ``EMC-effect'' region, and $x>0.6$, where nucleons' Fermi
motion is assumed to dominate   
 \cite{Deshpande:2005wd}.  
In Fig.\ref{fig1} we show the ratio, $R_A$, of the DIS nuclear structure
function over the free nucleon one for ${\rm ^{12}C}$. 
We focus on the region $x>0.1$, {\it i.e.} we omit shadowing because the 
very different space-time characteristics of this phenomenon 
would deserve a separate and lengthier discussion. 
%
There is currently no definitive consensus 
on the physics underlying nuclear effects at $x>0.1$.
It is by now clear that the so-called Light Cone (LC) convolution 
approach (see \cite{Roberts} for a review) is inadequate to 
reproduce the effect. However, the picture is emerging that corrections to 
this approximation given in terms
of the partons' transverse degrees of freedom, $k_\perp$, and off-shellness,
which are related to parton interactions might play an important role.   
In Ref.\cite{SL1}, within Hard Scattering Factorization \cite{Ellis:1982cd}, 
it was shown that  ``active-$k_\perp$'' 
effects on one side enhance the nuclear binding 
correction to the structure function, and on the other 
they are responsible for antishadowing. 
Inclusion of these effects produces the full curve in Fig.\ref{fig1}. 

We close this section by noting that while some of the puzzles concerning
{\it e.g.} the role of particles off-shellness and/or re-interactions in nuclei
will be addressed by dedicated inclusive scattering 
experiments {\it e.g.} in the Jefferson Lab program at 12 GeV
\cite{12GeV}, distinctively new type of information will also emerge 
from a class of exclusive type experiments measuring GPDs in nuclei, currently
being considered.

\section{Generalized Parton Distributions in Nuclei}
In this Section we illustrate the new type of information 
on the DIS structure of nuclei that can be obtained from GPDs.
The formalism for calculating GPDs in spin zero nuclei is described in 
\cite{SL1}. The amplitude for DVCS is a generalization of the forward virtual
Compton scattering one, whose imaginary part yields the DIS cross section, 
to the ``off-forward'' case namely when there is a momentum difference, $\Delta$ 
between the incoming and outgoing photon.
In case of unpolarized scattering from a spin zero nucleus, it is parametrized in terms 
of a GPD, $H_A$ which depends on two off-forward extra variables, denoted as $\zeta=\Delta^+/P^+$,
$P$ being the target's momemtum, and $t= - \Delta^2$. 

\begin{figure}[t]
  \vspace{6.0cm}
  \includegraphics{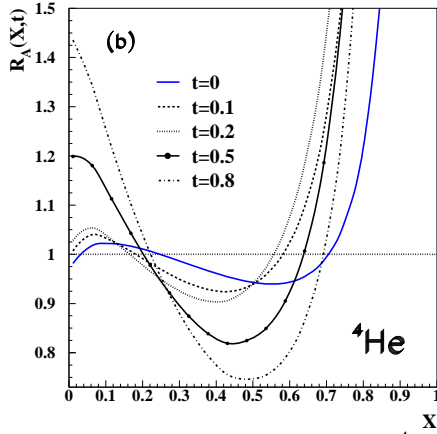}
  \caption{\it Ratio of nuclear GPD in $^4He$ to the free nucleon one for 
different values of $t$. }
    \label{fig2}
\end{figure}

In Fig.\ref{fig2} we show a ratio similar to the DIS ratio of Fig.1, of the nuclear GPD
over the nucleon one, normalized by their respective form factors. All curves were 
obtained at $\zeta=0$, namely the momentum difference is only transverse. 
%
The full curve in the figure
corresponds to the forward case. One can notice an enhancement of both the antishadowing and 
EMC-effect regions with increasing $t$. In simple terms this is due to the fact that transverse
d.o.f. and parton re-interactions are emphasized in a nucleus with respect to the free nucleon, and
GPDs are a more sensitive probe of such components than in the forward, inclusive 
scattering case. Technically, the enhancement 
with respect to the forward case is obtained
because nuclear GPDs are described in terms of $t$-dependent nuclear 
binding and average transverse momenta contributing to particles' off-shellness. One is
therefore dealing withr a sort of binding and transverse momentum ``form factors''
rather than the given, fixed experimental values determining the forward DIS case (Fig.1). 
 
In conclusion, the graphs in Fig\ref{fig2} illustrate a simple example of the many possibilities 
for a new insight into the deep inelastic structure 
of nuclei offered by considering a new class of exclusive experiments including DVCS.
We have shown in particular that parton interactions/off-shellness that have been advocated
as one of the possible explanations of the differences between the bound and free nucleons
structure functions, are enhanced in the off-forward case. Many other aspects including the
transverse spatial structure of nuclei and the connection with in medium
properties of the nucleon \cite{SL1}, are currently also being studied.

This work was completed under the US Department of Energy grant no.
DE-FG02-01ER41200.


%
\end{document}